\documentstyle[11pt,epsfig]{article}
\newcommand \be{\begin{equation}}
\newcommand \ba{\begin{eqnarray}}
\newcommand \ee{\end{equation}}
\newcommand \ea{\end{eqnarray}}
\newcommand{\lp}{\left(}
\newcommand{\rp}{\right)}

\begin{document}

\title{Characterization of large price variations in financial markets}
\thispagestyle{empty}

\author{Anders Johansen\footnote{Present address: Wind Energy Department, 
Ris\o \ National Laboratory, P.O. Box 49, DK-4000 Roskilde, Denmark. email:
anders.johansen@risoe.dk} \\
Niels Bohr Institute \\
Blegdamsvej 17, DK-2100 Kbh. \O  \\ DENMARK} 
\date{\today}
\maketitle

\abstract{
Statistics of drawdowns (loss from the last local maximum to the next local
minimum) plays an important role in risk assessment of investment strategies. 
As they incorporate higher ($>$ two) order correlations, they offer a better
measure of real market risks than the variance or other cumulants of daily (or 
some other fixed time scale) of returns. Previous results have shown that the 
vast majority of drawdowns occurring on the major financial markets have a 
distribution which is well-represented by a stretched exponential, while the 
largest drawdowns are occurring with a significantly larger rate than predicted
by the bulk of the distribution and should thus be characterized as {\it 
outliers} \cite{outl1,outl2}. In the present analysis, the definition of 
drawdowns is generalised to coarse-grained drawdowns or so-called 
$\epsilon$-drawdowns and a link between such {\it $\epsilon$-outliers} and 
preceding log-periodic power law bubbles previously identified \cite{SJ2001} 
is established.}

\thispagestyle{empty}
\pagenumbering{arabic}
\newpage
\setcounter{page}{1}

\section{Introduction} 

The characterization of stock market moves and especially large drops, 
{\it i.e.}, large negative moves in the price, are of profound importance
to risk management. A drawdown is defined as a persistent decrease in the 
price over consecutive days. A drawdown is thus the cumulative loss from the 
last maximum to the next minimum of the price, specifically the daily close
in the analysis presented here. Since the definition of 
``maximum'' and ``minimum'' is not unique except in a strict mathematical 
sense a drawdown may be defined in slightly varying ways. The definition used 
in the present paper is as follows. A drawdown is defined as the relative 
decrease in the price from a local maximum to the next local minimum {\it 
ignoring} price increases in between the two of maximum (relative or absolute)
size $\epsilon$. We will refer to this definition of drawdowns as 
``$\epsilon$-drawdowns'', where we will refer to $\epsilon$ as the 
{\it threshold}.

Drawdowns embody a rather subtle dependence since they are constructed from 
runs of the same sign variations. Their distribution thus captures the way 
successive drops can influence each other and construct in this way a 
(quasi-)persistent process. This persistence is not measured by the 
distribution of returns because, by its very definition, it forgets about
the relative positions of the returns as they unravel themselves as a function
of time by only counting their frequency\footnote{Realising this allows one
to construct synthetic price data with the same return distribution by a 
reshuffling of the returns\protect\cite{JS2000}}. This is not detected either by
the two-point correlation function, which measures an {\it average} linear 
dependence over the whole time series and as a consequence wash out features
only occurring at special times.

Related to the characterisation of drawdowns in the financial markets is the
concept of ``outliers'' \cite{outl1,outl2}. An outlier to some specific 
distribution may defined as a point which position deviates sufficiently from
those of the bulk of the distribution, it ``lies out'', as to arise suspicion 
that different processes are responsible for the generation of one hand the
overall distribution and on the other hand the outlier. Actually, testing for 
``outliers'' or more generally for a change of population in a distribution is 
a quite subtle problem. This subtle point is that the evidence for outliers and
extreme events does not require and is not even synonymous in general with the 
existence of a break in the distribution of the drawdowns. An example of this
comes from the distribution for the square of the velocity variations in 
shell models of turbulence. Naively, one would expect that the same physics 
apply in each shell layer (each scale) and, as a consequence, the distributions
in each shell should be the same, up to a change of unit reflecting the 
different scale embodied by each layer. The remarkable conclusion of L'vov
{\it et al.} \cite{L'vov} is that the distributions of velocity increment seem 
to be composed of two regions, a region of so-called ``normal scaling'' and a 
domain of extreme events, the ``outliers''.

Other groups have recently presented supporting evidence that crash and rally 
days significantly differ in their statistical properties from the typical 
market days. Lillo and Mantegna investigated the return distributions of an 
ensemble of stocks simultaneously traded in the New York Stock Exchange (NYSE) 
during market days of extreme crash or rally in the period from January 1987 
to December 1998 \cite{Lillo}. Out of two hundred distributions of 
returns, one for each of two hundred trading days where the ensemble of returns
is constructed over the whole set of stocks traded on the NYSE, anomalous large
widths and fat tails are observed specifically on the day of the crash of Oct. 
19 1987, as well as during a few other turbulent days. Specifically, they show 
that the overall shape of the distributions is modified in crash and rally 
days. Closer to our claim that markets develop precursory signatures of bubbles
of long time scales, Mansilla \cite{Mansilla} has also shown, using a measure 
of relative complexity, that time sequences corresponding to ``critical''
periods before large market corrections or crashes have more novel informations
with respect to the whole price time series than those sequences corresponding 
to periods where nothing happened. The conclusion is that the intervals where 
no financial turbulence is observed, that is, where the markets works fine,
the informational contents of the price time series is small.  In contrast, 
there seems to be significant information in the price time series associated 
with bubbles.

In a series of papers, the authors 
\cite{SJ2001,JS2000,SJB,SJ1997,JS1999,JSL1999,JLS2000} 
have shown that on the FX and major
stock markets, crashes are often preceded by precursory characteristics
quantified by a log-periodic power law, specifically
\be \label{lpeq}
p\lp t\rp = A+B\lp t_c - t\rp^z +C\lp t_c - t\rp^z \cos\lp  \omega 
\ln\lp t_c - t\rp - \phi \rp
\ee
The results on the US stock markets have been confirmed by several independent 
groups \cite{feigen,vandewalle,wolfgangpaul}

\begin{figure}
\begin{center}
\parbox[l]{8.5cm}{
\epsfig{file=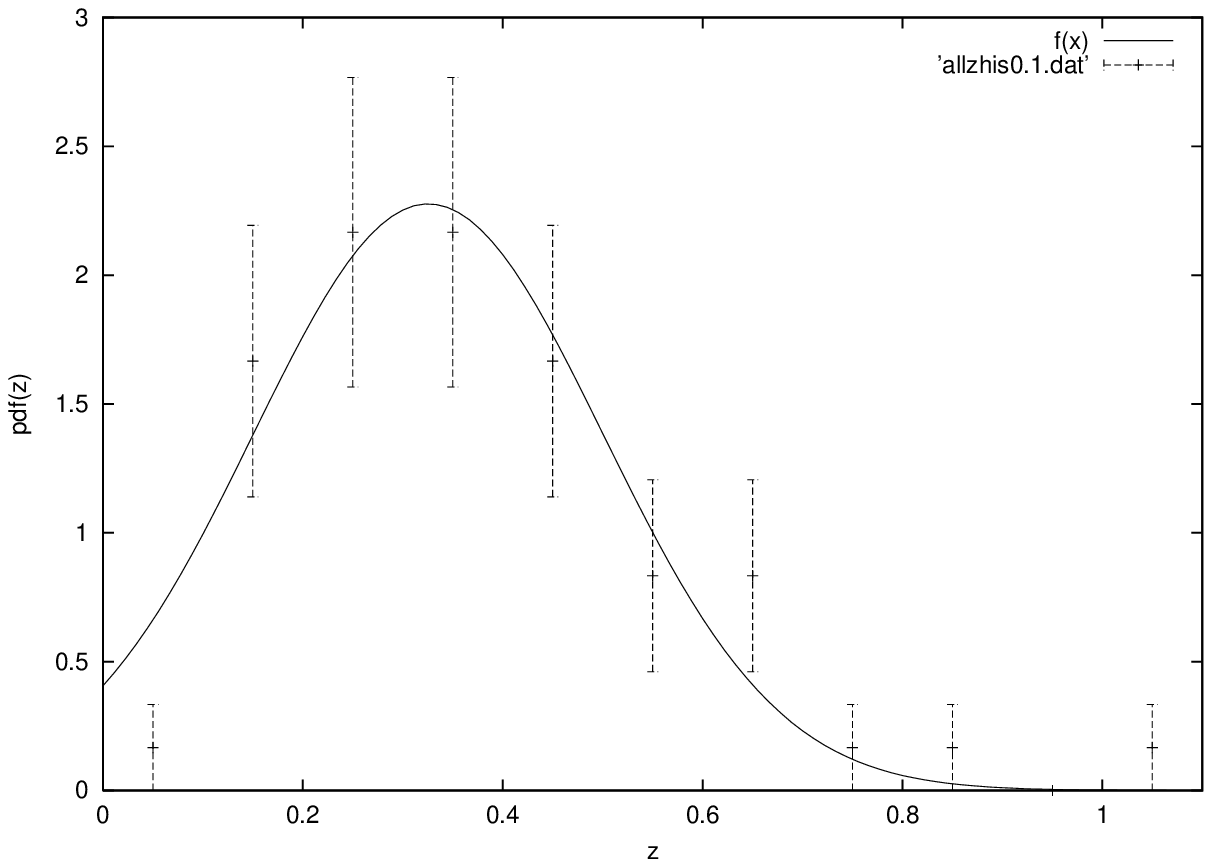,height=8cm,width=8.5cm}
\caption{\label{z} Distribution of fitted exponents $z$ in eq. 
(\protect\ref{lpeq} for over 30 bubbles. The fit is a Gaussian 
with mean $0.33$ and standard deviation $0.18$} }
\hspace{5mm}
\parbox[r]{8.5cm}{
\epsfig{file=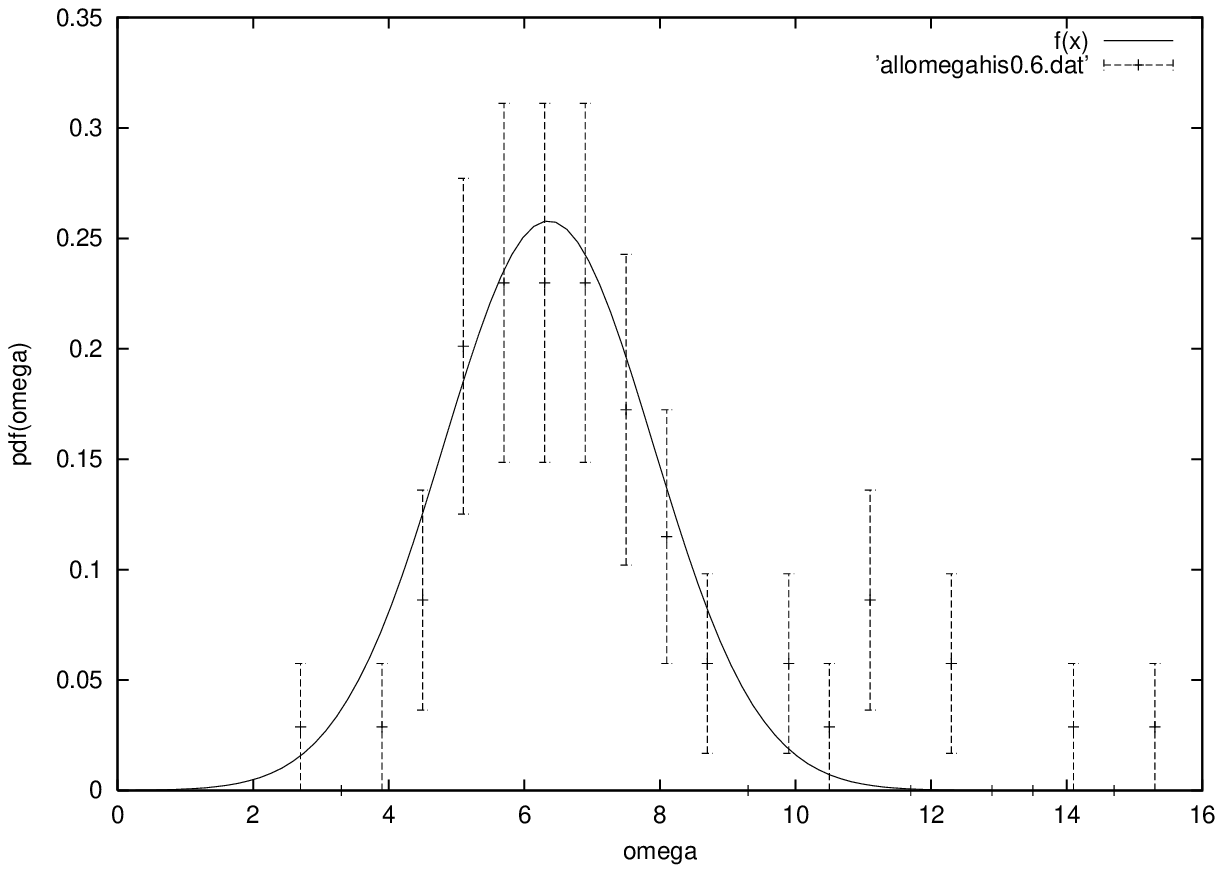,height=8cm,width=8.5cm}
\caption{\label{omega}Distribution of fitted log-frequencies $\omega$ in eq. 
(\protect\ref{lpeq} for over 30 bubbles. The fit is a Gaussian with 
mean $6.36$ and standard deviation $1.55$ } }

\vspace{1.5cm}

\parbox[l]{8.5cm}{
\epsfig{file=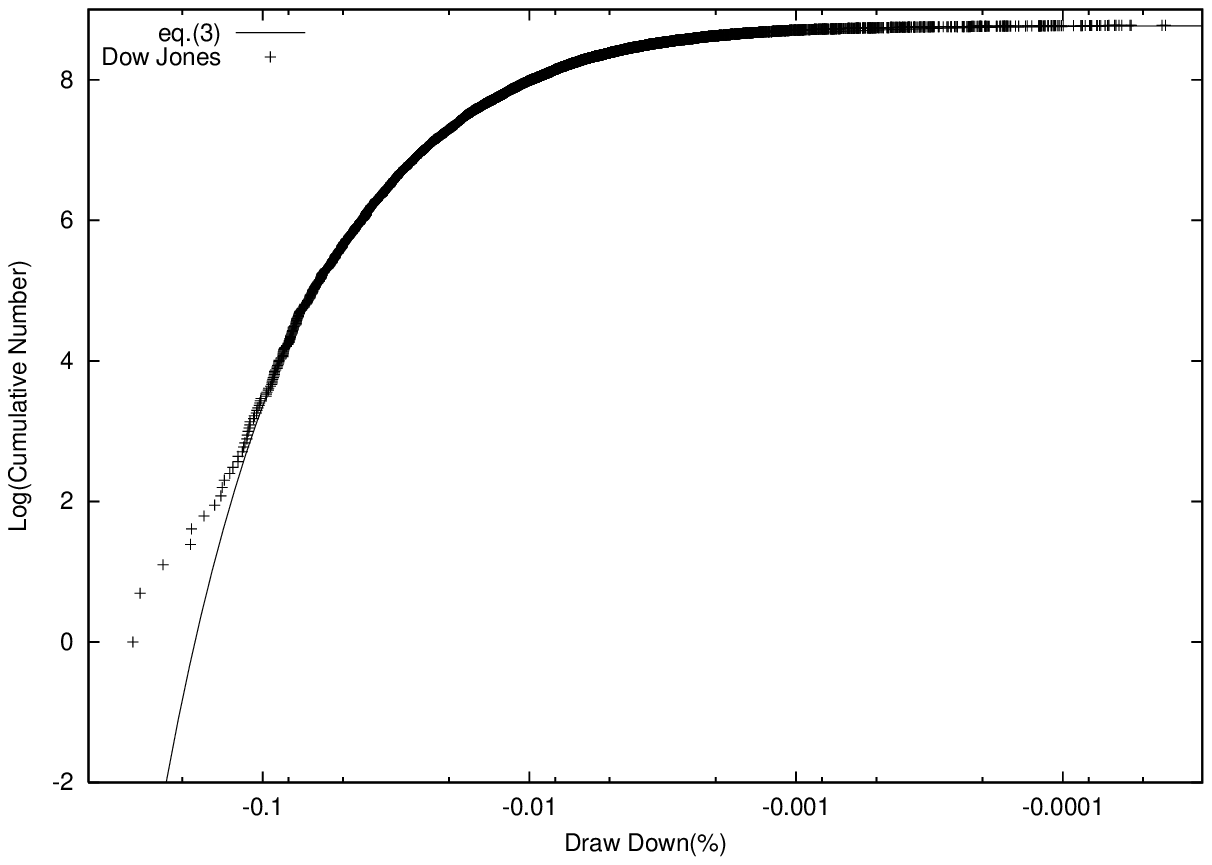,height=8cm,width=8.5cm}
\caption{\label{djdd}Natural logarithm of the cumulative distribution of 
drawdowns in the DJIA since 1900 until 2 May 2000. The fit is 
$\ln(N)=\ln(6469)-36.3 x^{0.83}$, where $6469$ is the total number of 
drawdowns.} }
\parbox[r]{8.5cm}{
\epsfig{file=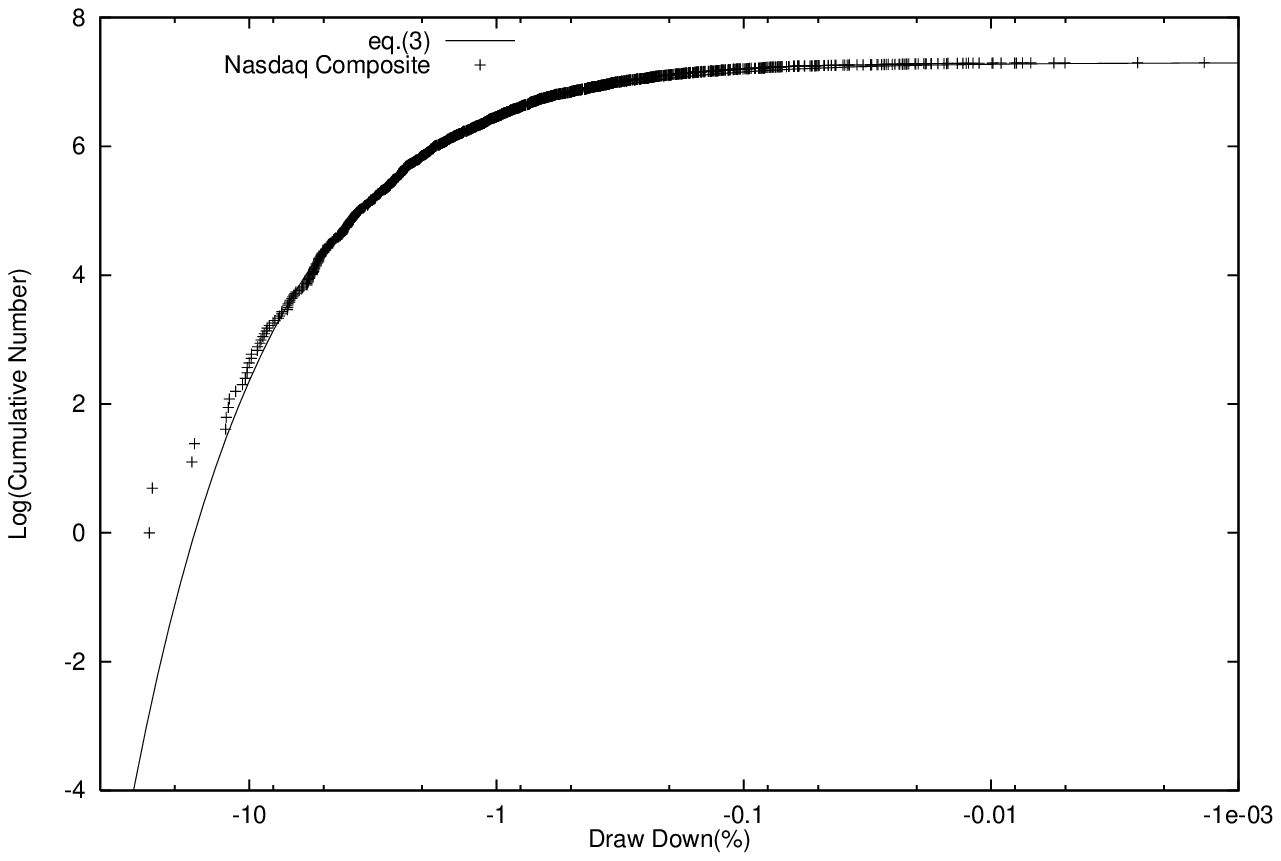,height=8cm,width=8.5cm}
\caption{\label{nasdd}Natural logarithm of the cumulative distribution of 
drawdowns in the Nasdaq since its establishment in 1971 until 18 April 2000. 
The fit is $\ln(N) = \ln(1479)-29.0 x^{0.77}$, where $1479$ is the total 
number of draw downs.} }
\end{center}
\end{figure}

Eq. (\ref{lpeq}) has its origin in a Landau-expansion type of argument and the 
underlying {\it Scaling Ansatz} is simply
\be \label{ansatz}
\frac{d F\lp x\rp}{d\ln x}=\alpha F\lp x\rp +z  F^2\lp x\rp \ldots
\ee
which to first order leads to eq. (\ref{lpeq}) with an arbitrary choice of
periodic function. The concept that only relative changes are important has a 
solid foundation in finance, but a detailed and rigorous derivation or 
justification for the Ansatz (\ref{ansatz}) has not been achieved. Instead the 
predictions that comes from applying this Ansatz, specifically those related 
to $t_c$, $z$ and $\omega$, to the data has been compared using different 
markets and time periods. Specifically, we have found that 
$$
\omega \approx 6.36 \pm 1.56 \hspace{10mm} z \approx 0.33 \pm 0.18
$$
for over thirty crashes on the major financial markets, see figures \ref{z} 
and \ref{omega}. This and the analysis to be presented in the following leads 
to a consistent and coherent picture when combined with the outlier concept.
A comment on figure \ref{omega} is necessary here. A fit with eq. (\ref{lpeq}) 
will often generate more than one solution. In general, the best fit in terms
of the r.m.s. is also the most sensible solution in terms of estimating the
{\it physical} variables $z$ and $\omega$ as well as the most probable time
$t_c$ of the end of the bubble, see \cite{SJ2001} for a more detailed
discussion. However, for a few cases the two best fits are included in the
statistics which explains the presence of the ``second harmonics'' around
$\omega \approx 11.5$.   

\section{Statistics and identification of bubbles and drawdowns }

Lately, an increasing amount of evidence that the largest negative market 
moves belongs to a different population than the smaller has accumulated 
\cite{outl1,outl2,SJ2001}. Specifically, it was found that the cumulative 
distributions of drawdowns on the worlds major financial markets, {\it e.g.},
the U.S. stock markets, the Hong-Kong stock market, the currency exchange
market (FX) and the Gold market are well parameterised by a stretched 
exponential
\be \label{stretched}
N(x) = Ae^{-bx^z} 
\ee
{\it except} for the $1\%$ (or less) largest drawdowns. In general, it was
found that the exponent $z \approx 0.8-0.9$ \cite{outl2}, see figures 
\ref{djdd} and \ref{nasdd} for two examples. It is worth noting that only the 
distributions
for the DJIA, the US\$/DM exchange rate and the Gold price exhibits clear 
outliers for the complement drawup distribution, whereas (all?) other markets
shows a strong asymmetry between the tails of the drawdown and drawup 
distributions the latter having no outliers. The range of the exponent for 
the drawup distributions is also generally higher with $z \approx 0.9-1.05$ 
except for the FX and Gold markets \cite{outl2}.

In the previous analysis and identification outliers on the major financial
markets \cite{outl2} drawdowns (drawups) were simply defined 
as a continuous decrease (increase) in the closing value of the price. Hence, 
a drawdown (drawup) was terminated by any increase (decrease) in the price no 
matter how small. A rather natural question concerns the effect of thresholding
on the distribution of drawdowns (drawups). There are two straightforward ways
to define a thresholded drawdown (drawup): We may ignore increases (decreases)
of a certain fixed magnitude (absolute or relative to the price) or we may 
ignore increases (decreases) over a fixed time horizon in both cases letting 
the drawdown (drawup) continue. We will refer this in general as coarse-grained
drawdowns (drawups) due to the smoothing obtained by ignoring small-scale 
fluctuations. In the present paper only the former definition of coarse-grained
drawdowns and drawups will be applied in the analysis, the latter being 
considered elsewhere \cite{epsilon}.

Price coarse-grained drawdowns can be defined as follows. We identify a local 
maximum in the price and look for a continuation of the downward trend ignoring
movements in the reverse direction smaller than $\epsilon$. Here $\epsilon$ 
is referred to as ``the threshold'', absolute or relative. Specifically, when 
a movement in the reverse direction is identified, the drawdown is nevertheless
continued {\it if } the magnitude, absolute or relative, is less than the
threshold. A very few drawdowns initiated by this algorithm end up as drawups 
and are discarded.

\begin{figure}
\begin{center}
\parbox[l]{8.5cm}{
\epsfig{file=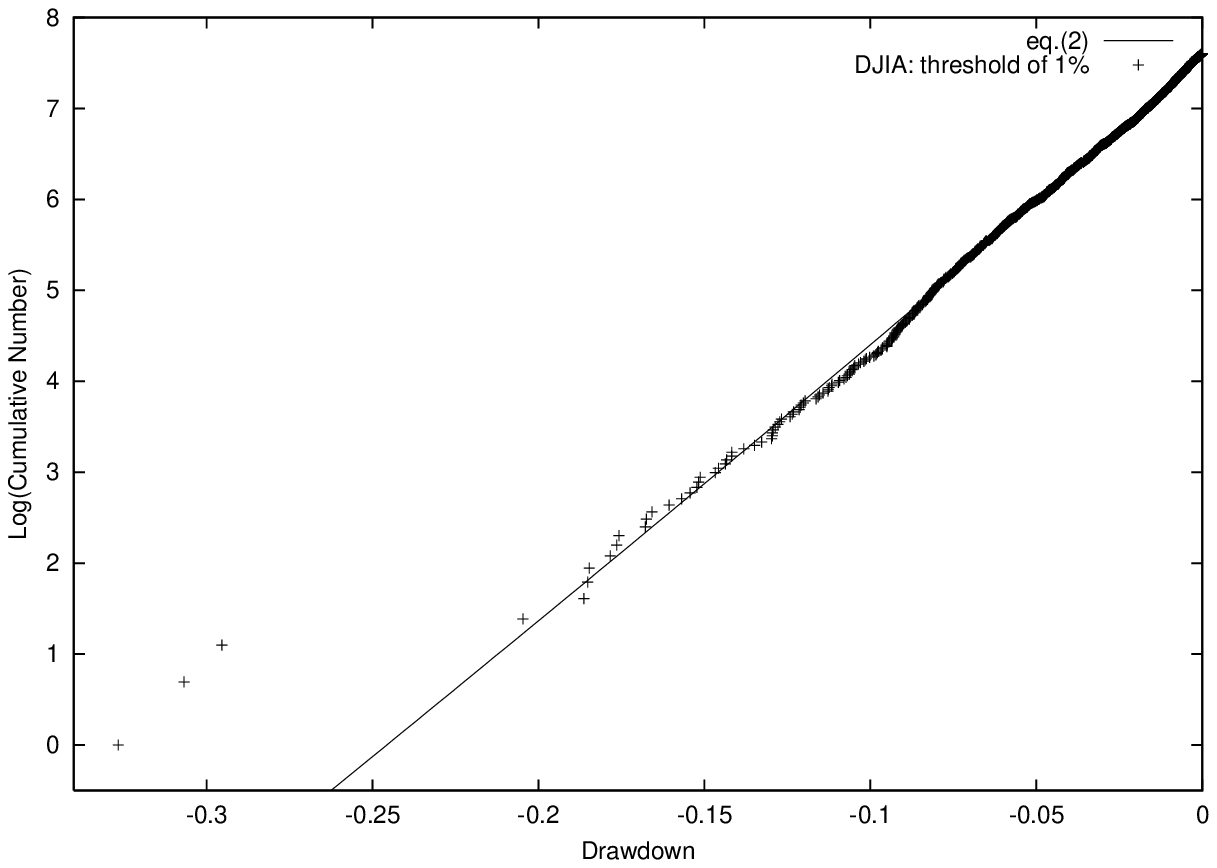,height=8cm,width=8.5cm}
\caption{\label{dj0.010ddthresh}DJIA 2004 events. Natural logarithm of the 
cumulative distribution of drawdowns coarse-grained with a relative threshold 
of $0.01$. The fit is $\ln(N) \approx 7.60 - 29.4 x^{0.96}$. } }
\hspace{5mm}
\parbox[r]{8.5cm}{
\epsfig{file=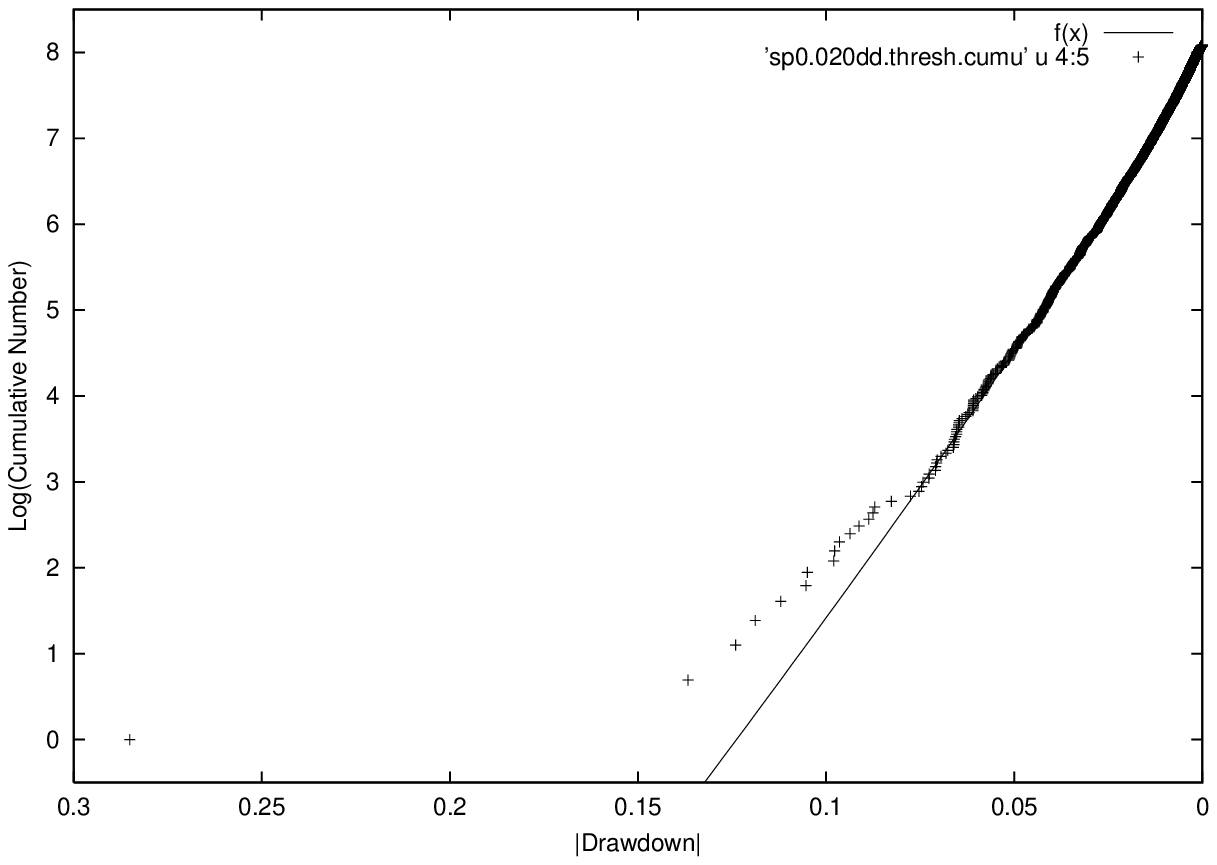,height=8cm,width=8.5cm}
\caption{\label{sp0.020ddthresh}SP500 3239 events. Natural logarithm of the 
cumulative distribution of drawdowns coarse-grained with an absolute 
threshold of $0.02$. The fit is $\ln(N) \approx  8.08 - 53.3 x^{0.90}$.} }
\vspace{1.5cm}

\parbox[l]{8.5cm}{
\epsfig{file=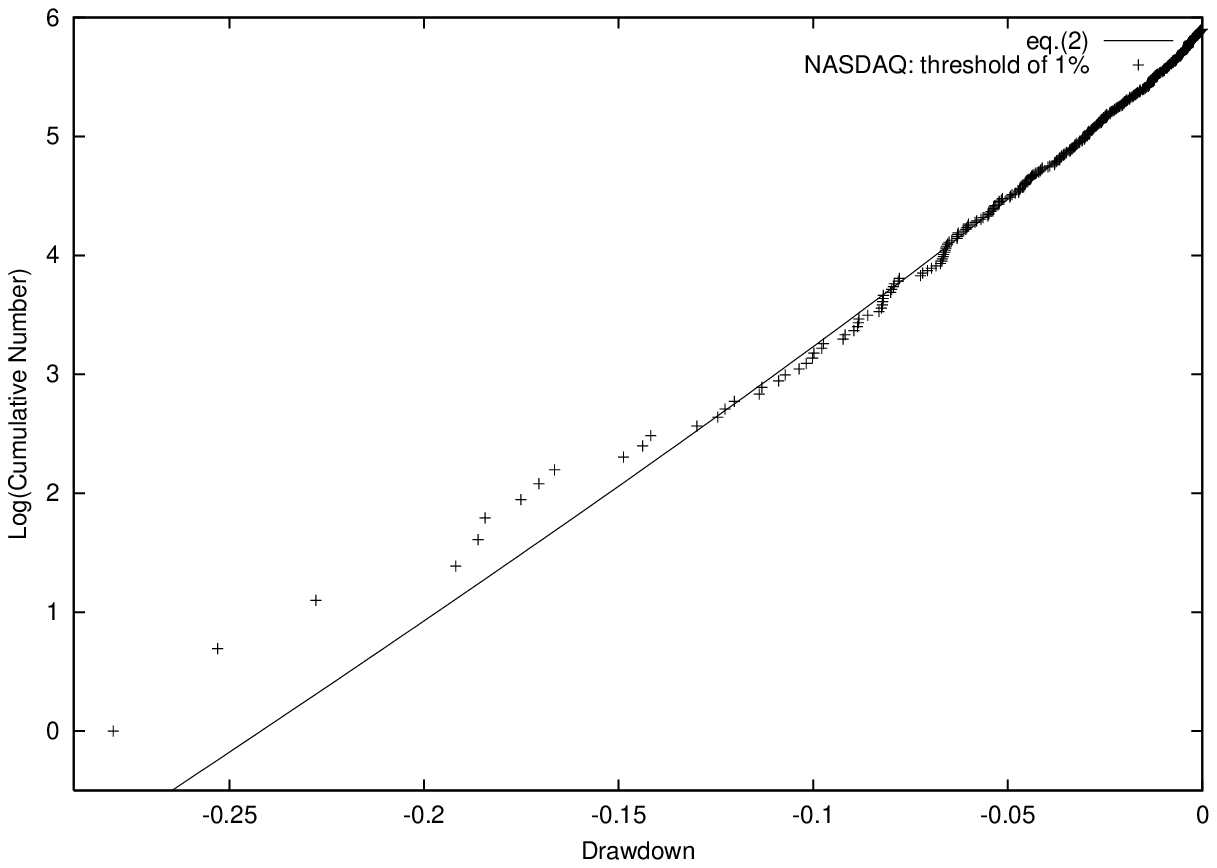,height=8cm,width=8.5cm}
\caption{\label{nas0.010ddthresh}Nasdaq 366 events. Natural logarithm of the 
cumulative distribution of drawdowns coarse-grained with a relative threshold 
of $0.01$. The fit is $\ln(N) \approx 5.91 - 21.1 x^{0.90}$} }
\hspace{5mm}
\parbox[r]{8.5cm}{
}
\end{center}
\end{figure}

In figures \ref{dj0.010ddthresh} to \ref{nas0.010ddthresh} we see the 
cumulative $\epsilon$-drawdown distributions of the DJIA, SP500 and Nasdaq.
The thresholds used were a relative threshold of $\epsilon =0.01$ for DJIA
and Nasdaq and an absolute threshold of $\epsilon =0.02$ for SP500, which
illustrates the problem of an objective determination of what threshold to use.
This problem will be addressed in more detail in a future publication elsewhere
\cite{epsilon}. We see that the fits with eq. (\ref{stretched}) for all three 
index fully captures the distributions {\it except} for a few cases which can 
be referred to as outliers. The dates of these outliers are
\begin{itemize}
\item DJIA: 1914, 1987 and 1929
\item SP500: 1987, 1962, 1998, 1987, 1974, 1946, 2000
\item Nasdaq: 1987, 2000, 1980, 1998, 1998, 1973, 1978,1987,1974
\end{itemize}

Except for the crashes related to the outbreak of WWI in 1914, the Yom Kibbur 
(Arab-Israeli) war and subsequent OPEC oil embargo in 1973 and the resignation 
and quite controversial pardoning of president R. Nixon in 1974, quite 
remarkably all of these outliers have log-periodic power law precursors 
well-described by eq. (\ref{lpeq}) and all, except for the Nasdaq crashes of 
1978 and 1980, have previously been published \cite{SJ2001}. The Nasdaq 
crashes of 1978 and 1980 was prior to this analysis unknown to us as the 
``pure'', {\it i.e.}, no threshold, presented in \cite{outl1,outl2} did not 
reveal its ``outlier nature''. 

The results presented here means that the {\it joint evidence} from the 
distributions of drawdowns in the DJIA, the SP500 and the Nasdaq identifies 
all crashes with log-periodic power law precursors found on the US stock market 
except the crash of 1937\footnote{Increasing the threshold for the DJIA does 
not improve this miss. Using a temporal coarse-graining as defined previously 
with a delay of $>2$ days attributes a drawdown of $\approx 19\%$ to the crash 
of 1937 but still places in (the far end of) the bulk of the distribution.}. 
Using $\epsilon = 0.02$ for the Nasdaq reduces the number of obvious outliers 
to five, two related to the 1987 and 2000 crashes each and one to the 1998 
crash. Using $\epsilon = 0.02$ for the DJIA changes nothing whereas this
threshold will remove all outliers except the crash of 1987 as outliers 
and, excluding the historical events of 1914, 1973 and 1974, vice versa.

Naturally, the optimal threshold (according to some specific definition) used 
in the outlier identification 
process is related to that particular index volatility. However, the volatility
is again nothing but a measure of the two-point correlations present in the 
index, which we have proven to be a insufficient measure when dealing with 
extreme market events. Hence, the thresholding procedure proposed and used 
here is a preliminary step toward more sophisticated amplification tools 
designed to better capture higher order correlations responsible for the 
extreme market events \cite{epsilon}.

\section{Conclusion}

The analysis presented here have strengthen the evidence for outliers in
the financial markets and that the concept can be used as a objective and
quantitative definition of a market crash. Furthermore, we have shown that
the existence of outliers in the drawdown distribution is primarily related 
to the existence of log-periodic power law bubbles prior to the occurrence of 
these outliers or crashes. In fact, of the 19 large drawdowns identified as 
outliers only 3 did not have prior log-periodic power law bubble and these
3 outliers could be linked to a specific major historical event. In complement,
only 1 (1937) previously identified log-periodic power law bubble was not 
identified as an outlier.

Further work is needed to clarify the role of different coarse-graining 
methods as well as to arrive at an objective choice for $\epsilon$. Last, 
any microscopic market model of log-periodic power law bubbles followed by 
large crashes should be address the question of why is the mean for 
$\omega = 2\pi / \ln \lambda$ so close to $2\pi$ giving a value for the 
discrete scaling factor $\lambda \approx e$.

\newpage

Cited papers by the authors are available from 
http://www.nbi.dk/\~\/johansen/pub.html.

\newpage

\end{document}